\documentclass[preprintnumbers,amsmath,amssymb,floatfix,11pt,prd,onecolumn,superscriptaddress,nofootinbib]{revtex4}
\usepackage{hyperref}
\usepackage{footnote}
\usepackage{amssymb}
\usepackage{amsmath}
\usepackage{graphicx}
\usepackage{dcolumn}
\usepackage{bm}
\usepackage{color}
\usepackage{xcolor}
\usepackage{float}
\usepackage{lipsum}
\allowdisplaybreaks
\usepackage{subfig}
\usepackage{comment}
\usepackage{multirow}
\usepackage[figurename=FIGURE,tablename=TABLE]{caption}
\usepackage{threeparttable}
\usepackage{hyperref}
\hypersetup{
    colorlinks=true, 
    linktoc=all,     
    linkcolor=magenta,
    citecolor=red
}

\newcommand{\ba}{\begin{eqnarray}}
\newcommand{\ea}{\end{eqnarray}}

\begin{document}

\title{Constraining $\beta$-Exponential Inflation with the latest ACT observations}

\author{Jureeporn Yuennan}
\email{jureeporn\_yue@nstru.ac.th}
\affiliation{Faculty of Science and Technology, Nakhon Si Thammarat Rajabhat University, Nakhon Si Thammarat, 80280, Thailand}

\author{Farruh Atamurotov}
\email{atamurotov@yahoo.com}
\affiliation{University of Tashkent for Applied Sciences, Str. Gavhar 1, Tashkent 100149, Uzbekistan}
\affiliation{Research Center of Astrophysics and Cosmology, Khazar University, 41 Mehseti Street, Baku AZ1096, Azerbaijan}
\affiliation{Tashkent State Technical University, Tashkent 100095, Uzbekistan}

\author{Salvatore~Capozziello}
\email{capozziello@na.infn.it}
\affiliation{Dipartimento di Fisica ``E. Pancini", Universit\`a di Napoli ``Federico II", Complesso Universitario di Monte Sant’ Angelo, Edificio G, Via Cinthia, I-80126, Napoli, Italy,}
\affiliation{Istituto Nazionale di Fisica Nucleare (INFN), sez. di Napoli, Via Cinthia 9, I-80126 Napoli, Italy,}
\affiliation{Scuola Superiore Meridionale, Largo S. Marcellino, I-80138, Napoli, Italy.}

\author{Phongpichit Channuie}
\email{phongpichit.ch@mail.wu.ac.th}
\affiliation{School of Science \& College of Graduate Studies, Walailak University, Nakhon Si Thammarat, 80160, Thailand}

\date{\today}

\begin{abstract}
Recent observations from the Atacama Cosmology Telescope (ACT), especially when combined with DESI baryon acoustic oscillation data, indicate a scalar spectral index \( n_s \) higher than the value reported by \textit{Planck} 2018, placing tension on universal inflationary attractor models. Motivated by this discrepancy, we investigate the inflationary predictions of the \(\beta\)-exponential potential, $
V(\phi)=V_0\left(1-\lambda\beta\phi/M_p\right)^{1/\beta}$
considering both minimally and non-minimally coupled realizations. This potential generalizes standard exponential inflation and naturally arises in braneworld scenarios. We derive analytical expressions for the slow-roll parameters and inflationary observables using a perturbative expansion in the non-minimal coupling \(\xi\), and validate these results through numerical calculations. In the minimally coupled case, the model predicts
\( n_s \simeq 0.976 \) and \( r \simeq 0.035 \) for \(N=50\) and moderate values of \(\beta\), remaining compatible with ACT+DESI (P-ACT-LB) constraints at the \(1\sigma\) level while yielding a spectral tilt larger than the universal attractor prediction. Introducing a small non-minimal coupling significantly improves agreement with observations by suppressing the tensor-to-scalar ratio while preserving the enhanced scalar tilt. For \(N=60 \), \( \lambda \sim 0.3\!-\!0.5 \), and \( \beta \sim \mathcal{O}(1\!-\!5) \), the non-minimally coupled model yields
\( n_s \simeq 0.974\!-\!0.976 \) and \( r \lesssim 0.03 \), comfortably consistent with ACT, DESI, and BICEP/Keck bounds. Our results show that the $\beta$-exponential potential, especially when implemented with a non-minimal coupling, exhibits good agreement with the latest CMB observations. Our inflationary predictions of the non-minimal model of $n_s$ and $r$ confirming the leading-order contributions in $\xi$ are sufficient to capture the essential features of both $r$ and $n_s$ in observationally relevant regimes.
\end{abstract}


\maketitle

\newpage
\section{Introduction}
Recent analyses combining the Atacama Cosmology Telescope (ACT) observations~\cite{ACT:2025fju,ACT:2025tim} with the Dark Energy Spectroscopic Instrument (DESI) survey data~\cite{DESI:2024uvr,DESI:2024mwx} have prompted the cosmology community to revisit the standard inflationary paradigm.
The latest ACT findings reveal that the scalar spectral index of primordial curvature perturbations deviates by about $2\sigma$ from the {\it Planck} 2018 result~\cite{Planck:2018jri}, suggesting the need for refinements in the conventional inflationary framework.
Inflation continues to serve as a cornerstone of modern cosmology, elegantly resolving the flatness, horizon, and monopole problems of the Big Bang model. Moreover, it provides a natural explanation for the origin of primordial fluctuations that gave rise to large-scale cosmic structures and manifest today as anisotropies in the cosmic microwave background (CMB)~\cite{Starobinsky:1980te,Sato:1981qmu,Guth:1980zm,Linde:1981mu,Albrecht:1982wi}.
These perturbations are usually characterized by two key observables: the scalar spectral index $n_s$, which captures the scale dependence of scalar perturbations, and the tensor-to-scalar ratio $r$, describing the relative amplitude of primordial gravitational waves.
In canonical inflationary scenarios, both $n_s$ and $r$ are expressed in terms of the number of $e$-folds, $N$, between horizon exit and the end of inflation, allowing for a direct comparison with cosmological data.

A notable and widely applicable prediction common to many models is the so-called “universal attractor” relation, $n_s = 1 - \tfrac{2}{N}$, found in a broad class of models such as $\alpha$-attractors~\cite{Kallosh:2013tua,Kallosh:2013hoa,Kallosh:2013maa,Kallosh:2013yoa,Kallosh:2014rga,Kallosh:2015lwa,Roest:2015qya,Linde:2016uec,Terada:2016nqg,Ueno:2016dim,Odintsov:2016vzz,Akrami:2017cir,Dimopoulos:2017zvq}, the Starobinsky $R^2$ model~\cite{Starobinsky:1980te}, and Higgs inflation involving a large non-minimal coupling to gravity~\cite{Kaiser:1994vs,Bezrukov:2007ep,Bezrukov:2008ej}.
The work in Ref.~\cite{Chang:2019ebx} examined one-loop finite-temperature corrections to curvature perturbations in Higgs inflation and demonstrated that thermal effects improve agreement with {\it Planck} CMB data.
Similar predictions are also obtained in composite inflaton models~\cite{Karwan:2013iph,Channuie:2012bv,Bezrukov:2011mv,Channuie:2011rq}, as summarized in the reviews~\cite{Channuie:2014ysa,Samart:2022pza}.
For the typical benchmark $N=60$, the universal attractor relation yields $n_s \approx 0.9667$, consistent with the {\it Planck} 2018 measurement of $n_s = 0.9649 \pm 0.0042$~\cite{Planck:2018jri}.
However, the latest ACT data~\cite{ACT:2025fju,ACT:2025tim} indicate a slightly higher spectral index.
The combined {\it Planck}–ACT (P-ACT) analysis gives $n_s = 0.9709 \pm 0.0038$, while including CMB lensing and BAO information from DESI (P-ACT-LB) raises it further to $n_s = 0.9743 \pm 0.0034$. As regards the running index $\alpha_s$ of the scalar tilt $n_s$, we use the observational constraints \cite{AtacamaCosmologyTelescope:2025nti}, $\alpha_s = 0.0062 \pm 0.0104$, i.e., the $\alpha_s$ is between $-0.004$ and $+0.017$ at 95\% C.L. These updated results create significant tension with the universal attractor class, excluding them at approximately the $2\sigma$ level and even challenging the Starobinsky $R^2$ model itself~\cite{ACT:2025fju}.
In particular, Ref.~\cite{Gialamas:2025kef} examined Higgs-like inflation with radiative corrections under the light of the ACT observations.
This unexpected outcome has spurred extensive theoretical efforts to reconcile inflationary models with the new ACT results~\cite{Kallosh:2025rni,Gao:2025onc,Liu:2025qca,Yogesh:2025wak,Yi:2025dms,Peng:2025bws,Yin:2025rrs,Byrnes:2025kit,Wolf:2025ecy,Aoki:2025wld,Gao:2025viy,Zahoor:2025nuq,Ferreira:2025lrd,Mohammadi:2025gbu,Choudhury:2025vso,Odintsov:2025eiv,Odintsov:2025wai,Q:2025ycf,Zhu:2025twm,Kouniatalis:2025orn,Hai:2025wvs,Dioguardi:2025vci,Yuennan:2025kde,Oikonomou:2025xms,Oikonomou:2025htz,Odintsov:2025jky,Aoki:2025ywt,Gialamas:2025kef,Gialamas:2025ofz,Yuennan:2025tyx,Pallis:2025nrv,Pallis:2025gii,Yuennan:2025inm,Yuennan:2025mlg,Nojiri:2026hij,Oikonomou:2026gae}, with a comprehensive summary presented in~\cite{Kallosh:2025ijd}.

In this work, we explore both minimally and non-minimally coupled inflationary models in which the early Universe’s accelerated expansion is governed by a class of potentials that extend the traditional power-law inflation through a generalized exponential function~\cite{Alcaniz:2006nu}.
As highlighted in Ref.~\cite{Santos:2017alg}, the so-called $\beta$-exponential potential naturally arises within the framework of braneworld cosmology and provides a compelling fit to current observational constraints.
Our investigation has two main objectives: first, to update the analysis of the minimally coupled $\beta$-exponential model presented in Ref.~\cite{Santos:2017alg} using the most recent {\it Planck} 2018 results, and to introduce and examine the cosmological implications of the non-minimally coupled $\beta$-exponential scenario.

The structure of this paper is as follows.
In Section~(\ref{II}), we present a concise overview of theories with a scalar field non-minimally coupled to gravity and introduce the non-minimally coupled $\beta$-exponential inflationary model. This section also includes the derivation of the slow-roll parameters and the analytical expressions for the key inflationary observables, namely the scalar spectral index $n_s$ and the tensor-to-scalar ratio $r$, considering both minimal and non-minimal coupling cases. We also perform a numerical analysis of the inflationary dynamics to compare with the perturbative expansions in $\xi$ in Section~(\ref{Sec5}). In Section~(\ref{sec4}), we confront the theoretical predictions with the most recent observational constraints from the Atacama Cosmology Telescope (ACT) and DESI collaborations. Finally, the main conclusions and findings of this work are summarized in Section~(\ref{Con}).

\section{Slow-roll Analysis}\label{II}
In this section, we revisit the formalism of a scalar field non-minimally coupled to gravity, 
outlining the conditions under which slow-roll inflation occurs and deriving the key observable quantities. 
In the Jordan frame, the general action describing a non-minimally coupled scalar field is given by
\begin{eqnarray}
S=\int d^4x\sqrt{-g}\Bigg[ \frac{\Omega^2(\phi)}{2\kappa^2}R 
- \frac{1}{2}g^{\mu\nu}\partial_{\mu}\phi\partial_{\nu}\phi 
- V(\phi) \Bigg],
\label{1}
\end{eqnarray}
where $R$ is the Ricci scalar, and $\Omega^2(\phi)\equiv 1+\kappa^2\xi\phi^2$. 
Here, $\xi$ denotes the coupling strength between the scalar field and gravity, $\kappa^2 = 8\pi G = M_p^{-2}$ with $M_p$ being the reduced Planck mass, and $V(\phi)$ represents the inflationary potential. 
Evidently, for $\xi=0$, the action reduces to the standard minimally coupled case. 

The presence of the coupling function $\Omega^2(\phi)$ renders the field equations rather intricate. 
To express them in a more familiar Einstein-like form, we perform a conformal transformation of the metric, defined as
\begin{gather}
g_{\mu\nu}\longrightarrow\hat{g}_{\mu\nu}=\Omega^2 g_{\mu\nu}.
\label{2}
\end{gather}
Under this transformation, the Ricci scalar can be re-expressed in terms of the new metric, 
leading to the action in the Einstein frame,
\begin{eqnarray}
S=\int d^4x\sqrt{-g}\Bigg[
\frac{1}{2\kappa^2}R 
- \frac{1}{2}F^2(\phi)\hat{g}^{\mu\nu}\partial_{\mu}\phi\partial_{\nu}\phi - U(\phi)
\Bigg],
\label{3}
\end{eqnarray}
where
\begin{gather}
F^2(\phi)\equiv\frac{1+\kappa^2\xi\phi^2(1+6\xi)}{(1+\kappa^2\xi\phi^2)^2}, 
\label{4}\\
U(\phi)\equiv\frac{V(\phi)}{(1+\kappa^2\xi\phi^2)^2}.
\label{5}
\end{gather}
Introducing a canonical scalar field $\chi$ through the field redefinition
\begin{eqnarray}
\chi_{\phi}\equiv\frac{d\chi}{d\phi}
=\sqrt{\frac{1+\kappa^2\xi\phi^2(1+6\xi)}{(1+\kappa^2\xi\phi^2)^2}},
\label{6}
\end{eqnarray}
the action becomes
\begin{eqnarray}
S=\int d^4x\sqrt{-g}
\Bigg[\frac{1}{2\kappa^2}R
-\frac{1}{2}g^{\mu\nu}\partial_{\mu}\chi\partial_{\nu}\chi
-U(\chi)\Bigg],
\label{7}
\end{eqnarray}
which corresponds to the Einstein-frame representation of Eq.~(\ref{1}). Assuming the slow-roll approximation, where the inflaton evolves gradually along the potential, 
the first-two slow-roll parameters are defined (for the potential in Eq.~\ref{5}) as
\begin{eqnarray}
\epsilon = \frac{M_p^2}{2}\left( \frac{V_{\phi}}{V \chi_{\phi}}\right)^2, 
\quad 
\eta = M_p^2\left(\frac{V_{\phi\phi}}{V \chi_{\phi}^2} 
- \frac{V_{\phi}\chi_{\phi\phi}}{V\chi_{\phi}^3}\right)\,.
\label{8}
\end{eqnarray}
Inflation proceeds as long as $\epsilon, |\eta| \ll 1$, 
and terminates once one of these parameters reaches unity, e.g., $\epsilon(\phi_{\text{end}})=1$. The key inflationary observables are the scalar spectral index $n_s$, 
which quantifies the scale dependence of the curvature perturbations; 
the tensor-to-scalar ratio $r$, representing the relative amplitude of tensor modes. They are expressed in terms of the slow-roll parameters as~\cite{Lyth:1998xn}
\begin{eqnarray}
n_s = 1 - 6\epsilon + 2\eta, 
\quad 
r = 16\epsilon\,.
\label{9}
\end{eqnarray}
The amplitude of scalar perturbations at the pivot scale $k_{\star}$ can be written as
\begin{gather}
P_{R\star} = \frac{1}{24\pi^2 M_p^4}
\frac{U(\phi_{\star})}{\epsilon(\phi_{\star})}\Big|_{k=k_{\star}},
\label{10}
\end{gather}
where $\phi_{\star}$ is the field value at horizon crossing. We adopt $P_{R} = 2.0933\times 10^{-9}$ at $k_{\star} = 0.05~\text{Mpc}^{-1}$, 
consistent with {\it Planck} 2018 data~\cite{Planck:2018jri}. 
The same analysis yields a scalar spectral index 
$n_s = 0.9649 \pm 0.0042$ (68\% C.L.) 
and an upper bound on the tensor-to-scalar ratio, 
$r < 0.056$, from the combined {\it Planck} and BICEP2/Keck Array observations~\cite{BICEP:2021xfz}. To compare the minimally and non-minimally coupled models, we start by examining the minimal coupling case.
\begin{figure}
\includegraphics[width=8 cm]{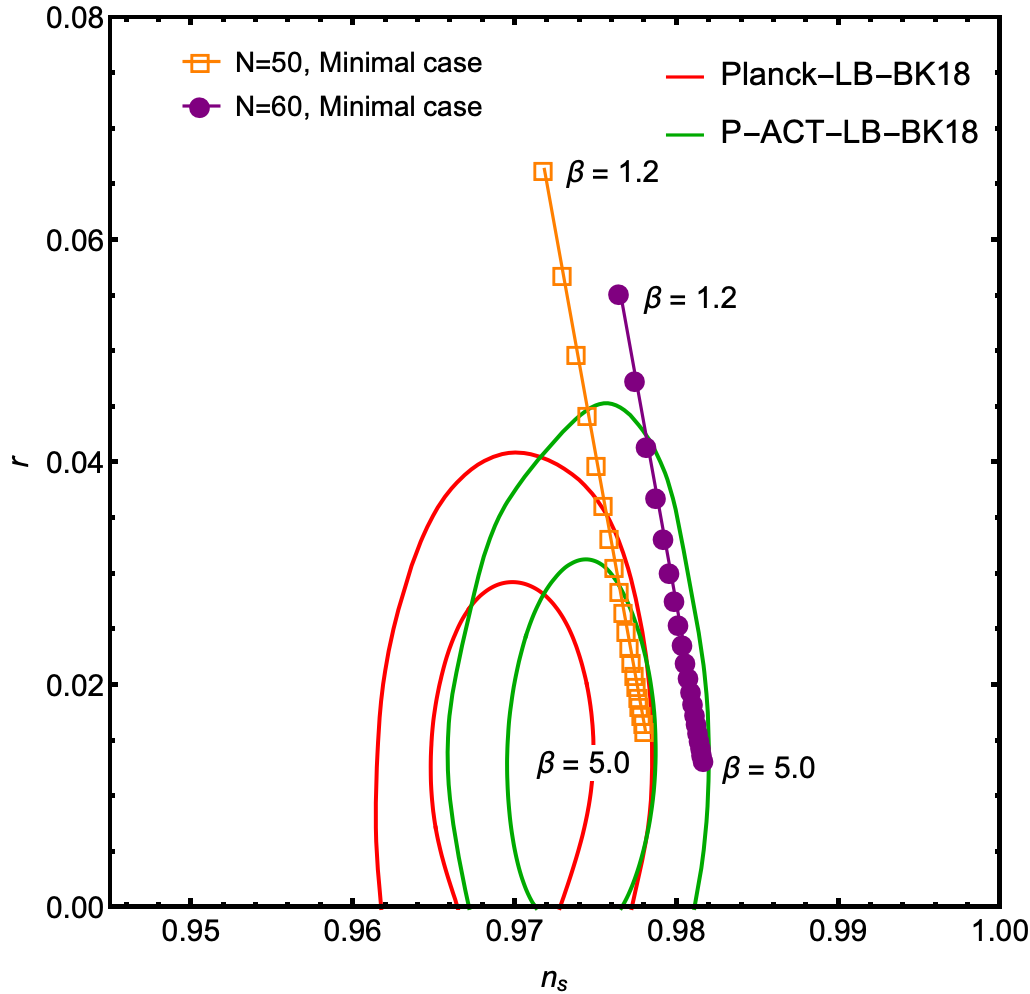}
\caption{Constraints on the scalar and tensor primordial power spectra, shown in the $r-n_{s}$
parameter space. The bounds on $r$ are primarily determined by the BK18 observations, whereas the limits on $n_s$ are set by Planck (red) and P-ACT (green) data. We fix $N=50,\,60$ and vary a parameter $\beta$ for a minimally-coupled model.}\label{rns}
\end{figure}

\subsection{Minimally coupled case}
Introduced in Ref.~\cite{Alcaniz:2006nu}, this model serves as a phenomenological framework extending the standard power-law inflation. The inflationary potential considered in this work is expressed as
\begin{gather}
V(\phi) = V_{0}\left( 1 - \lambda\beta\frac{\phi}{M_p} \right)^{\frac{1}{\beta}},
\label{12}
\end{gather}
which was later motivated within the braneworld framework in Ref.~\cite{Santos:2017alg}. 
From a mathematical perspective, Eq.~(\ref{12}) represents a generalized form of the exponential potential, parameterized by $\beta$, 
such that in the limit $\beta \rightarrow 0$, the standard exponential potential is recovered. 
The potential is defined to be positive, $V>0$, when $1-\lambda\beta\phi/M_p>0$, and vanishes otherwise, i.e. $V=0$ for $1-\lambda\beta\phi/M_p\leq0$. Defining $x\equiv \phi/M_{p}$, the slow-roll parameters take the form
\begin{gather}
\epsilon = \frac{\lambda^2}{2(1-\lambda\beta \,x)^2}, 
\quad 
\eta = \frac{\lambda^2(1-\beta)}{(1-\lambda\beta\,x)^2},
\label{13}
\end{gather}
and the field value at the end of inflation, $x_{\text{end}}$, is obtained by imposing the condition $\epsilon=1$, yielding
\begin{gather}
x_{\text{end}} =x^{\pm}= \frac{1}{\beta}\left(\frac{1}{\lambda}\pm\frac{1}{\sqrt{2}}\right)\,,
\label{14}
\end{gather}
for which the latter solution is viable $x_{\text{end}} =x^{-}$ \cite{dosSantos:2021vis}. The positivity of Eq.(\ref{14}) can be achieved when $\beta >0\land 0<\lambda <\sqrt{2}$. It is worth noting that Ref.~\cite{dosSantos:2021vis} has previously examined the observational viability of this class of models by performing a comprehensive Bayesian analysis in light of the most recent Cosmic Microwave Background (CMB) data from the Planck Collaboration. In contrast to their approach, the present study aims to obtain explicit analytical insights into the inflationary dynamics. Specifically, we derive closed-form expressions for the relevant observables by employing a perturbative technique in the regime of small non-minimal coupling $\xi$. This analytical treatment not only facilitates a clearer understanding of the model’s dependence on its parameters but also allows for a direct comparison between the minimally and non-minimally coupled cases, see Refs.\cite{Bostan:2025may,Bostan:2024ugi} also its incorporation with an $R^{2}$ term within the Palatini formulation. By integrating the number of $e$-folds,
\begin{gather}
N=\frac{1}{M^{2}_{p}}\int_{\phi_{\text{end}}}^{\phi_{\star}}\frac{V \chi^{2}_{\phi}}{V_{\phi}}\,d\phi,
\end{gather}
we find that the field value at horizon crossing reads
\begin{gather}
x_{\star} = \frac{1}{\beta\lambda}-\frac{1}{\beta}\sqrt{\frac{1}{2}+2\beta {\cal N}}\,,
\label{15}
\end{gather}
where $N_{\rm end}$ is defined as
\begin{eqnarray}
N_{\rm end}=\frac{\beta x^2_{\rm end}}{2}-\frac{x_{\rm end}}{\lambda }\,.
\label{6zz2}
\end{eqnarray}
The main inflationary observables, the scalar spectral index $n_s$ and the tensor-to-scalar ratio $r$, can
then be expressed as
\begin{eqnarray}
n_s&=&1 -6\epsilon + 2\eta 
\nonumber\\&=& \frac{\lambda ^2 \left(\beta  \left(\beta  x^2-2\right)-1\right)-2 \beta  \lambda  x+1}{(\beta  \lambda  x-1)^2}, \\
r &=& 16\epsilon = \frac{8 \lambda ^2}{(1-\beta  \lambda  x)^2}.
\label{16}
\end{eqnarray}
In terms of the e-folding number, $N$, they can be recast as
\begin{eqnarray}
n_s&=&1-\frac{2 (1+2 \beta)}{1+4 \beta N}, \label{ns1}\\
r &=& \frac{16}{1+4 \beta  N}.
\label{17}
\end{eqnarray}
It is worth noting that, from Eqs.~(\ref{ns1}) and (\ref{17}), 
the above relations can be expressed solely in terms of \(N\) \& \(\beta\), indicating that they are no longer dependent on \(\lambda\). For minimal model, we particularly display the predictions for the tensor-to-scalar ratio $r$ and the spectral index $n_s$ with $N=50$ and $N=60$ for various values of $\beta$ in Table (\ref{tab:ng3}).
\begin{table}[t]
\centering
	\begin{tabular}{c|c|c|c|c|c|c|c}
	\quad $N$ \quad\quad & \quad\quad$\beta$\quad\quad\quad & \quad\quad$r$ \quad\quad\quad & \quad\quad$n_s$\quad\quad\quad & \quad\quad$N$\quad\quad\quad & \quad\quad$\beta$\quad\quad\quad & \quad\quad$r$ \quad\quad\quad & \quad\quad$n_s$\quad\quad\quad\\\hline
    & $0.6$ & $0.1322$ & $0.9636$ &  & $0.6$ & $0.1103$
    & $0.9697$\\
    & $0.7$ & $0.1135$ & $0.9660$&   & $0.7$& $0.0947$ & $0.9716$ \\
    $50$ & $0.9$ & $0.0884$ & $0.9691$ & $60$ & $0.9$ & $0.0737$ & $0.9742$ \\
    & $3.3$ & $0.0242$ & $0.9770$ &    & $3.3$ & $0.0202$ & $0.9808$ \\
    & $5.0$ & $0.0160$ & $0.9780$ &    & $5.0$ & $0.0133$ & $0.9817$\\\hline
	\end{tabular}
	\caption{We present specific predictions for the tensor-to-scalar ratio $r$ and the scalar spectral index $n_s$ in the minimally coupled model, fixing $N=50$ and $N=60$ while varying $\beta$.} \label{tab:ng3}
\end{table}

\subsection{Non-minimally coupled case}
\begin{figure}
    \centering
    \includegraphics[width=0.8\linewidth]{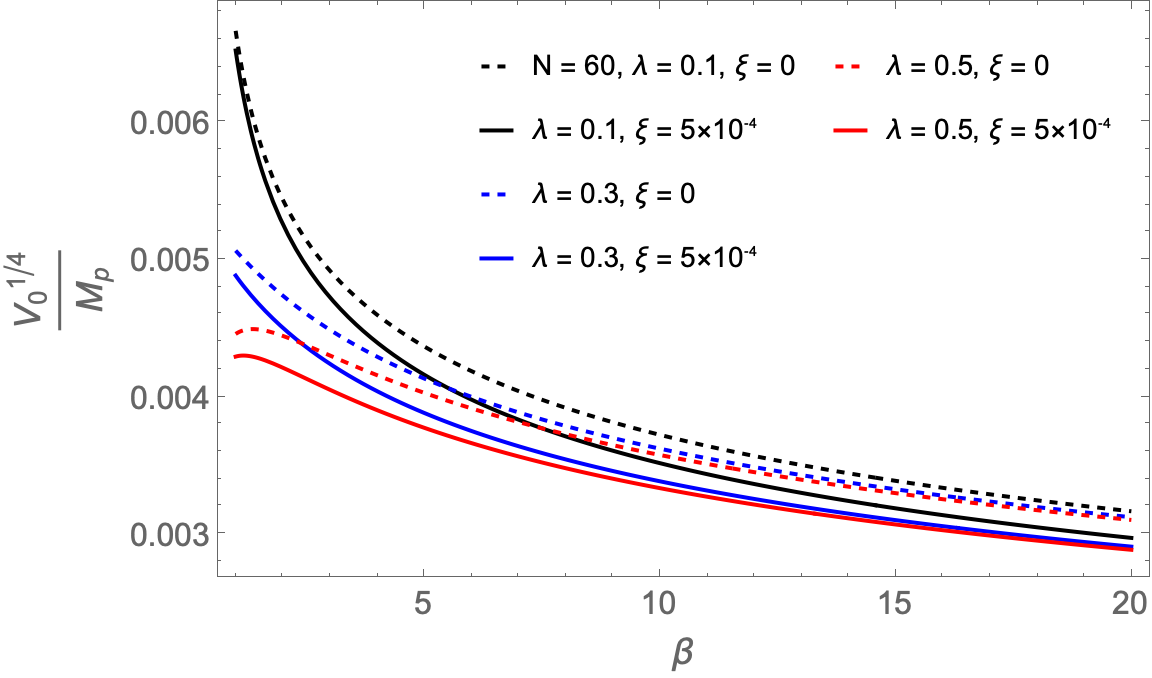}
    \caption{The variation of the normalized potential \( V_0^{1/4}/M_p \) as a function of \( \beta \) 
for different values of the coupling parameter \( \lambda \), with the number of e-folds fixed at \( N = 60 \). The black curves correspond to \( \lambda = 0.1 \), where the dashed line denotes the minimally coupled case (\( \xi = 0 \)) and the solid line represents the nonminimally coupled case (\( \xi = 0.0005 \)). The purple curves show the results for \( \lambda = 0.3 \), while the orange curves correspond to \( \lambda = 0.5 \); in both cases, dashed lines indicate minimal coupling (\( \xi = 0 \)) and solid lines indicate 
nonminimal coupling (\( \xi = 0.0005 \)).}\label{V01}
\end{figure}

In what follow, we consider next the nonminimally-coupled model. In the case of the non-minimally coupled \(\beta\)-exponential model, 
the potential in the Einstein frame takes the form
\begin{eqnarray}
U(\chi(\phi))\equiv\frac{V(\phi(\chi))}{(1+\kappa^2\xi\phi^{2}(\chi))^2}.
\label{pot}
\end{eqnarray}
where $\phi(\chi)$ is given by Eq.~(\ref{6}). Using the above potential, the first-two slow-roll parameters can be determined to obtain
\begin{eqnarray}
\epsilon &=& \frac{M_p^2}{2}\left( \frac{U_{\phi}}{U \chi_{\phi}}\right)^2=\frac{(\lambda +\xi  x (x (\lambda -4 \beta  \lambda )+4))^2}{2 \left(\xi  (6 \xi +1) x^2+1\right) (\beta  \lambda  x-1)^2}\,,\label{epp}\\
\eta &=& M_p^2\left(\frac{U_{\phi\phi}}{U \chi_{\phi}^2} 
- \frac{U_{\phi}\chi_{\phi\phi}}{U\chi_{\phi}^3}\right)\nonumber\\&=& \Big(\left(\xi  (6 \xi +1) x^2+1\right)^2 (\beta  \lambda  z-1)^2\Big)^{-1}\Bigg(\lambda ^2 \Big(4 \beta ^2 \xi  x^2 \left(\xi  x^2 \left(4 \xi  (6 \xi +1) x^2+3\right)-1\right)\nonumber\\&&-\beta  \left(\xi x^2+1\right) \left(\xi x^2 \left(4 \xi  \left(2 (6 \xi +1) x^2+3\right)+9\right)+1\right)+\left(\xi  x^2+1\right)^2 \left(\xi  (6 \xi +1) x^2+1\right)\Big)\nonumber\\&&+\lambda  \xi  x \left(\left(\xi  x^2+1\right) \left(\xi  \left(7 (6 \xi +1) x^2+6\right)+7\right)-8 \beta  \left(\xi  x^2 \left(4 \xi  (6 \xi +1) x^2+3\right)-1\right)\right)\nonumber\\&&+4 \xi  \left(\xi x^2 \left(4 \xi  (6 \xi +1) x^2+3\right)-1\right)\Bigg)\,.\label{etaa}
\end{eqnarray}
We may check that in the limit $\xi=0$ the expressions of $\epsilon$ and $\eta$ for the minimally
coupled case Eq.(\ref{13}) are simply recovered. The number of e-folds between the horizon exit of a given scale and the end of inflation, $N$, is in that case given by
\begin{eqnarray}
N&=&\int^{\chi_{\star}}_{\chi_{\rm end}}\frac{d\chi}{M_{p}}\frac{1}{\sqrt{2\epsilon}}=\frac{1}{M^{2}_{p}}\int^{\phi_{\star}}_{\phi_{\rm end}}\Bigg(\frac{U_{\phi}}{U\chi^{2}_{\phi}}\Bigg)^{-1}d\phi\nonumber\\&&=\frac{1}{M^{2}_{p}}\int^{x_{\star}}_{x_{\rm end}}\Bigg(\frac{\left(\xi  (6 \xi +1) x^2+1\right) (\beta  \lambda  x-1)}{\left(\xi  x^2+1\right) (\xi  x ((4 \beta -1) \lambda  x-4)-\lambda )}\Bigg)dx\,.\label{No}
\end{eqnarray}
It is, however, more convenient to express the equation in terms of the field $\phi$ and to explore the small-$\xi$ regime, as this allows one to verify the consistency of the results with the minimally coupled limit, $\xi = 0$. This assumption also facilitates the application of the perturbative approach. Using a perturbation method for (\ref{epp}) in the small parameters $\xi$ and searching
for a solution to this condition of the type:
\begin{eqnarray}
x_{\rm end}=y_{0}+y_{1}\xi, \quad{\rm with}\quad y_{0}=\frac{1}{\beta}\left(\frac{1}{\lambda}-\frac{1}{\sqrt{2}}\right)\,,
\end{eqnarray}
we end up with the solution for $\epsilon(\phi_{\rm end})=1$:
\begin{eqnarray}
x_{\rm end}=\frac{1}{\beta}\Bigg(\frac{1}{\lambda }-\frac{1}{\sqrt{2}}\Bigg)+\frac{1}{8 \beta ^4 \lambda ^4}\left(8 \sqrt{2} \beta ^2 \lambda ^4-16 \beta ^2 \lambda ^3-\sqrt{2} \beta  \lambda ^4+4 \beta  \lambda ^3-2 \sqrt{2} \beta  \lambda ^2\right)\xi+{\cal O}(\xi^{2}).
\end{eqnarray}
It should be noted that the first term represents the contribution from the minimally coupled case. From Eq. (\ref{No}), we find the expression up to first order in $\xi$:
\begin{eqnarray}
N=-\frac{\beta  \xi  \left(x_{\star}^4-x_{\rm end}^4\right)}{8}+\frac{\xi  \left(x_{\star}^3-x_{\rm end}^3\right)}{6 \lambda}+\frac{\beta  \left(x_{\star}^2-x_{\rm end}^2\right)}{2}-\frac{(x_{\star}-x_{\rm end})}{\lambda}\,.\label{num}
\end{eqnarray}
Inserting $x_{\rm end}$ into the above expression and using a perturbation method in the small parameters $\xi$ and search
for a solution to this condition of the type:
\begin{eqnarray}
x_{\star}=x_{0}+x_{1}\xi, \quad{\rm with}\quad x_{0}=\frac{1}{\beta  \lambda }-\frac{\sqrt{2+8 \beta {\cal N}}}{2 \beta }\,.
\end{eqnarray}
From (\ref{num}), the solution reads
\begin{eqnarray}
x_{\star}&=&\frac{1}{\beta  \lambda }-\frac{\sqrt{2+8 \beta N}}{2 \beta }\nonumber\\&&+\frac{1}{12 \beta ^3 \lambda ^2 \sqrt{8 \beta  N+2}}\Bigg(2 \sqrt{2} \beta  \lambda ^2+\beta  \Big(-16 \sqrt{2} \beta  \lambda ^2+3 \lambda ^3 (4 \beta  ((4 \beta -1) M (2 \beta  M+1)+2)-1)\nonumber\\&&-4 \sqrt{2} (2 \beta -1) \lambda ^2 (4 \beta  M+1)^{3/2}-6 \lambda  (4 \beta  M+1)\Big)\Bigg)\xi+{\cal O}(\xi^{2}).
\end{eqnarray}
Here we choose the solution $x_{\star}$ such that a condition $1-\lambda\,\beta \,x_{\star}>0$ is satisfied. Taking the value of $x_{\star}$ at horizon crossing, the slow-roll parameters $\epsilon$ and $\eta$ can be conveniently rewritten as functions of the e-folding number $N$:
\begin{eqnarray}
\epsilon_{\star}&=&\frac{1}{1+4 \beta N}\nonumber\\&&+\frac{1}{3 \beta ^3 \lambda ^3 (4 \beta N+1)^2}\Bigg(-8 \sqrt{2} \beta ^2 \lambda ^2+\sqrt{2} \beta  \lambda ^2-144 \beta ^4 \lambda ^3 N^2+12 \beta ^3 \lambda ^3 N^2\nonumber\\&&-72 \beta ^3 \lambda ^3 N+32 \sqrt{2} \beta ^3 \lambda ^2 N \sqrt{4 \beta N+1}+6 \beta ^2 \lambda ^3 N-4 \sqrt{2} \beta ^2 \lambda ^2 N \sqrt{4 \beta  N+1}\nonumber\\&&+8 \sqrt{2} \beta ^2 \lambda ^2 \sqrt{4 \beta N+1}+\left(-\sqrt{2}\right) \beta  \lambda ^2 \sqrt{4 \beta N+1}\Bigg)\xi+{\cal O}(\xi^{2})\,,
\end{eqnarray}
and
\begin{eqnarray}
\eta_{\star}&=&\frac{2 (1-\beta)}{4 \beta N+1}\nonumber\\&&+\frac{1}{3 \beta ^3 \lambda ^3 (4 \beta  N+1)^2}\Bigg(-36 \beta ^3 \lambda ^3+16 \sqrt{2} \beta ^2 \sqrt{\beta ^2 \lambda ^4}-18 \sqrt{2} \beta  \sqrt{\beta ^2 \lambda ^4}+2 \sqrt{2} \sqrt{\beta ^2 \lambda ^4}+3 \beta ^2 \lambda ^3\nonumber\\&&-288 \beta ^5 \lambda ^3 N^2-264 \beta ^4 \lambda ^3 N^2+24 \beta ^3 \lambda ^3 N^2-144 \beta ^4 \lambda ^3 N+32 \sqrt{2} \beta ^4 \lambda ^2 N \sqrt{4 \beta  N+1}\nonumber\\&&-132 \beta ^3 \lambda ^3 N+60 \sqrt{2} \beta ^3 \lambda ^2 N \sqrt{4 \beta  N+1}+8 \sqrt{2} \beta ^3 \lambda ^2 \sqrt{4 \beta  N+1}+12 \beta ^2 \lambda ^3 N\nonumber\\&&-8 \sqrt{2} \beta ^2 \lambda ^2 N \sqrt{4 \beta N+1}+15 \sqrt{2} \beta ^2 \lambda ^2 \sqrt{4 \beta  N+1}-2 \sqrt{2} \beta  \lambda ^2 \sqrt{4 \beta  N+1}\Bigg)\xi+{\cal O}(\xi^{2}).
\end{eqnarray}
Using the above solution for $x=x_{\star}$, the scalar spectral index $n_s$ and the tensor-to-scalar ratio $r$ can then be expressed as
\begin{eqnarray}
n_s&=&1-\frac{2 (1+2 \beta)}{1+4 \beta N}\nonumber\\&&+\frac{2}{3 \left(\beta ^3 \lambda ^3 (4 \beta  N+1)^2\right)}\Bigg(36 \beta ^3 \lambda ^3-16 \sqrt{2} \beta ^2 \sqrt{\beta ^2 \lambda ^4}-6 \sqrt{2} \beta  \sqrt{\beta ^2 \lambda ^4}+\sqrt{2} \sqrt{\beta ^2 \lambda ^4}-3 \beta ^2 \lambda ^3\nonumber\\&&+288 \beta ^5 \lambda ^3 N^2-168 \beta ^4 \lambda ^3 N^2+12 \beta ^3 \lambda ^3 N^2+144 \beta ^4 \lambda ^3 N-32 \sqrt{2} \beta ^4 \lambda ^2 N \sqrt{4 \beta  N+1}-84 \beta ^3 \lambda ^3 N\nonumber\\&&-8 \sqrt{2} \beta ^3 \lambda ^2 \sqrt{4 \beta  N+1}+36 \sqrt{2} \beta ^3 \lambda ^2 N \sqrt{4 \beta  N+1}+6 \beta ^2 \lambda ^3 N-4 \sqrt{2} \beta ^2 \lambda ^2 N \sqrt{4 \beta N+1}\nonumber\\&&+9 \sqrt{2} \beta ^2 \lambda ^2 \sqrt{4 \beta  N+1}+\left(-\sqrt{2}\right) \beta  \lambda ^2 \sqrt{4 \beta N+1}\Bigg)\xi+{\cal O}(\xi^{2})\,, \label{ns11}
\end{eqnarray}
and
\begin{eqnarray}
r &=& \frac{16}{4 \beta N+1}\nonumber\\&&+\frac{16}{3 \beta ^3 \lambda ^3 (4 \beta N+1)^2}\Bigg(-8 \sqrt{2} \beta  \sqrt{\beta ^2 \lambda ^4}+\sqrt{2} \sqrt{\beta ^2 \lambda ^4}-144 \beta ^4 \lambda ^3 N^2+12 \beta ^3 \lambda ^3 N^2\nonumber\\&&-72 \beta ^3 \lambda ^3 N+32 \sqrt{2} \beta ^3 \lambda ^2 N \sqrt{4 \beta  N+1}+6 \beta ^2 \lambda ^3 N-4 \sqrt{2} \beta ^2 \lambda ^2 N \sqrt{4 \beta  N+1}\nonumber\\&&+8 \sqrt{2} \beta ^2 \lambda ^2 \sqrt{4 \beta  N+1}-\sqrt{2} \beta  \lambda ^2 \sqrt{4 \beta  N+1}\Bigg)\xi+{\cal O}(\xi^{2}).
\label{171}
\end{eqnarray}
It is evident that, in the limit $\xi \rightarrow 0$, the results smoothly reduce to those of the minimally coupled model, confirming the consistency of our formulation with the minimal case. While higher-order terms in $\xi$ may influence the value of $r$, our analysis is restricted to first-order contributions in $\xi$, which are subsequently checked against numerical calculations.
\begin{table}[t]
\centering
\begin{tabular}{c|c|c|c|c|c|c|c|c}
\quad $N$ \quad\quad 
& \quad $\lambda$ \quad\quad
& \quad $\beta$ \quad\quad 
& \quad $r^{\rm ana}$ \quad\quad 
& \quad $r^{\rm num}$ \quad\quad & \,\, \% difference \quad
& \quad $n_s^{\rm ana}$ \quad\quad 
& \quad $n_s^{\rm num}$ \quad\quad & \,\, \% difference \quad\\
\hline
    &       & $1.00$ & $0.0500572$ & $0.0441718$ & $12.49\%$ & $0.972354$ & $0.971916$ & $0.045\%$ \\
    &       & $1.30$ & $0.0377917$ & $0.0339234$ & $10.79\%$ & $0.973690$ & $0.973031$ & $0.068\%$\\
$60$ & $0.5$ & $1.55$ & $0.0313322$ & $0.028461$ & $9.60\%$ & $0.974383$ & $0.973637$ & $0.077\%$\\
    &       & $3.35$ & $0.0139084$ & $0.0132721$ & $4.68\%$ & $0.976191$ & $0.975387$ & $0.082\%$\\
    &       & $4.85$ & $0.00946231$ & $0.00921378$ & $2.66\%$ & $0.976627$ & $0.975884$ & $0.076\%$\\
\hline
\end{tabular}
	\caption{We present specific predictions for the tensor-to-scalar ratio $r$ and the scalar spectral index $n_s$ in the non-minimally coupled model with $\xi=5\times 10^{-4}$, fixing $N=60$ and $\lambda=0.3$, while varying $\beta$.} \label{tab:non1}
\begin{tabular}{c|c|c|c|c|c|c|c|c}
\quad $N$ \quad\quad 
& \quad $\lambda$ \quad\quad
& \quad $\beta$ \quad\quad 
& \quad $r^{\rm ana}$ \quad\quad 
& \quad $r^{\rm num}$ \quad\quad & \,\, \% difference \quad
& \quad $n_s^{\rm ana}$ \quad\quad 
& \quad $n_s^{\rm num}$ \quad\quad & \,\, \% difference \quad\\
\hline
    &       & $1.00$ & $0.0477905$ & $0.0456223$ & $4.64\%$ & $0.972354$ & $0.971985$ & $0.038\%$ \\
    &       & $1.30$ & $0.0362111$ & $0.0349616$ & $3.51\%$ & $0.973630$ & $0.973145$ & $0.050\%$\\
$60$ & $0.5$ & $1.55$ & $0.0300970$ & $0.0292833$ & $2.74\%$ & $0.974298$ & $0.973769$ & $0.054\%$\\
    &       & $3.35$ & $0.0135009$ & $0.0135533$ & $0.39\%$ & $0.976072$ & $0.975534$ & $0.055\%$\\
    &       & $4.85$ & $0.0092256$ & $0.0093789$ & $1.64\%$ & $0.976513$ & $0.976019$ & $0.051\%$\\
\hline
\end{tabular}
	\caption{We present specific predictions for the tensor-to-scalar ratio $r$ and the scalar spectral index $n_s$ in the non-minimally coupled model with $\xi=5\times 10^{-4}$, fixing $N=60$ and $\lambda=0.5$, while varying $\beta$.} \label{tab:non2}
\end{table}
Considering the amplitude of scalar perturbations given in Eq.(\ref{10}), $V_{0}/M^{4}_{p}$ can be directly determined to obtain:
\begin{eqnarray}
\frac{V_{0}}{M_{p}} &\simeq& 2.21 \times 10^{-9} e^{0.346574/\beta } \left(\lambda  \sqrt{4\beta N+1}\right)^{-1/\beta }\Bigg(0.00422+0.0169 \beta  N\nonumber\\&&+\frac{0.000352}{\beta ^4 \lambda ^3}\Big(\beta \Big(45.255 (\beta +0.375) \sqrt{\beta ^2 \lambda ^4}+3 (12 \beta -1) \lambda ^3 (16 \beta  (\beta -0.5) N (\beta  N+0.5)-1)\nonumber\\&&-5.657 (\beta -1) (8 \beta -1) \lambda ^2 (4 \beta  N+1)^{3/2}-96 (\beta -0.25) \lambda  (\beta  N+0.25)\Big)\nonumber\\&&-2.828 \sqrt{\beta ^2 \lambda ^4}\Big)\xi\Bigg)^{-1}\,.\label{beta}
\end{eqnarray}
The behavior of $V_{0}$ is displayed in Fig.(\ref{V01}). In the limit $\xi \to 0$, the results smoothly approach those of the minimally-coupled case. For nonminimally-coupled model, we particularly report the predictions for the tensor-to-scalar ratio $r$ and the spectral index $n_s$ with $N=60$ for various values of $\beta$ in Table (\ref{tab:non1}). We find that the nonminimally-coupled model using $\lambda=0.5,\,\beta=1.55$ and $N=60$ predicts the tensor-to-scalar ratio $r=0.03$ and the spectral index $n_s=0.9743$ compatible with those reported by the observations.

\section{Numerical verification}\label{Sec5}
So far, all results for the nonminimally coupled model have been derived using perturbative expansions in $\xi$; however, we also perform a numerical analysis of the inflationary dynamics to compare with these two approaches. Since the predictions for $n_s$ and $r$ are highly sensitive to the field value at horizon crossing, it is necessary to numerically validate the analytical approximations to ensure their reliability across the parameter space used. Similarly to $x_{\rm end}$, the scalar field at the horizon crossing, $x_{\star}$, should be obtained numerically for values of $\beta, \lambda$ and $\xi$ satisfying Eq.(\ref{No}). Therefore, the scalar spectral index and the tensor-to-scalar ratio can be written as
\begin{eqnarray}
r=\frac{8(\lambda +\xi  x_{\star} (x_{\star} (\lambda -4 \beta  \lambda )+4))^2}{\left(\xi  (6 \xi +1) x_{\star}^2+1\right) (\beta  \lambda x_{\star}-1)^2}\,,\label{epp}\\
\end{eqnarray}
and
\begin{eqnarray}
n_{s} &=&1-\frac{3 (\lambda +\xi  x_{\star} (x_{\star} (\lambda -4 \beta  \lambda )+4))^2}{\left(\xi  (6 \xi +1) x^2+1\right) (\beta  \lambda  x_{\star}-1)^2}\nonumber\\&&+\frac{2}{\left(\xi  (6 \xi +1) x_{\star}^2+1\right)^2 (\beta  \lambda x_{\star}-1)^2}\Bigg(\Bigg(\lambda ^2 \Big(4 \beta ^2 \xi  x_{\star}^2 \left(\xi  x_{\star}^2 \left(4 \xi  (6 \xi +1) x_{\star}^2+3\right)-1\right)\nonumber\\&&-\beta  \left(\xi  x_{\star}^2+1\right) \left(\xi  z^2 \left(4 \xi  \left(2 (6 \xi +1) x_{\star}^2+3\right)+9\right)+1\right)+\left(\xi x_{\star}^2+1\right)^2 \left(\xi  (6 \xi +1) x_{\star}^2+1\right)\Big)\nonumber\\&&+\lambda  \xi  x_{\star} \left(\left(\xi x_{\star}^2+1\right) \left(\xi  \left(7 (6 \xi +1) x_{\star}^2+6\right)+7\right)-8 \beta  \left(\xi  x_{\star}^2 \left(4 \xi  (6 \xi +1) x_{\star}^2+3\right)-1\right)\right)\nonumber\\&&+4 \xi  \left(\xi  x_{\star}^2 \left(4 \xi  (6 \xi +1) x_{\star}^2+3\right)-1\right)\Bigg)\Bigg)\,,\label{epp}
\end{eqnarray}
from which one can verify that, in the limit $\xi\rightarrow 0$, the expressions for $n_s$ and $r$ reduce to those of the minimally coupled case given in Eq.(\ref{ns1}) and (\ref{17}).
\begin{figure}
\includegraphics[width=8 cm]{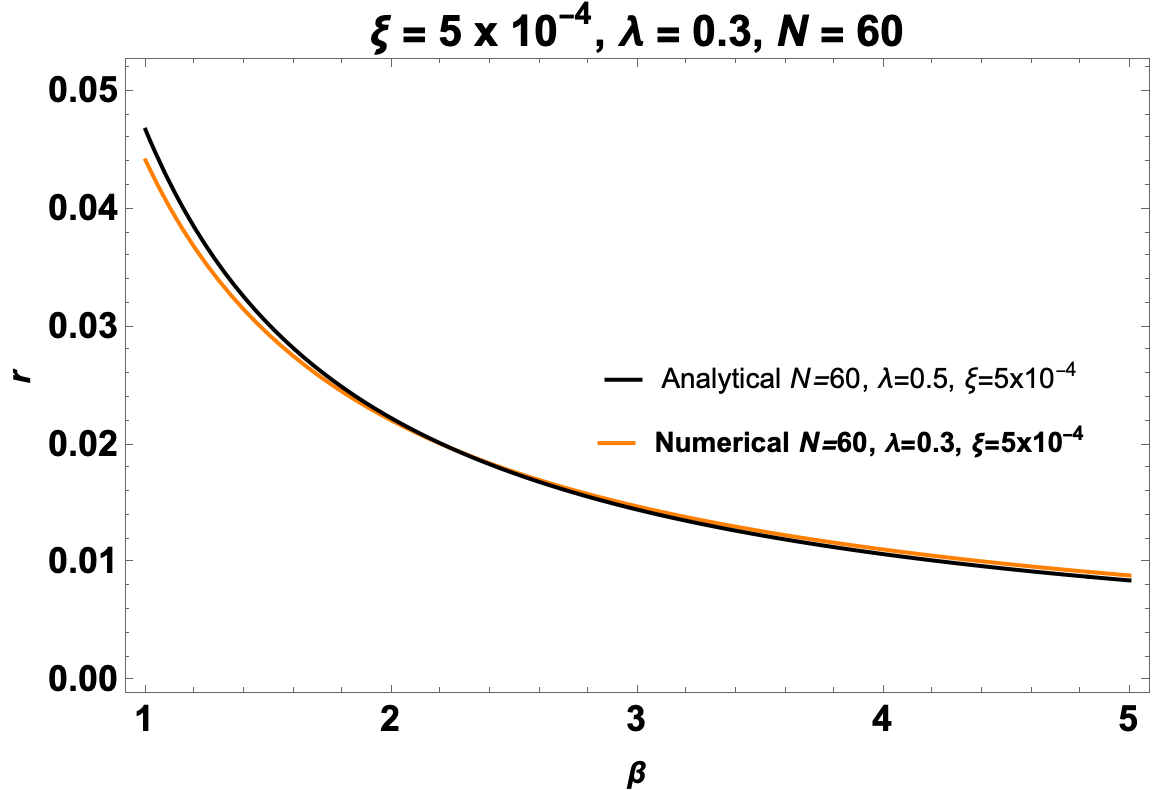}
\includegraphics[width=8 cm]{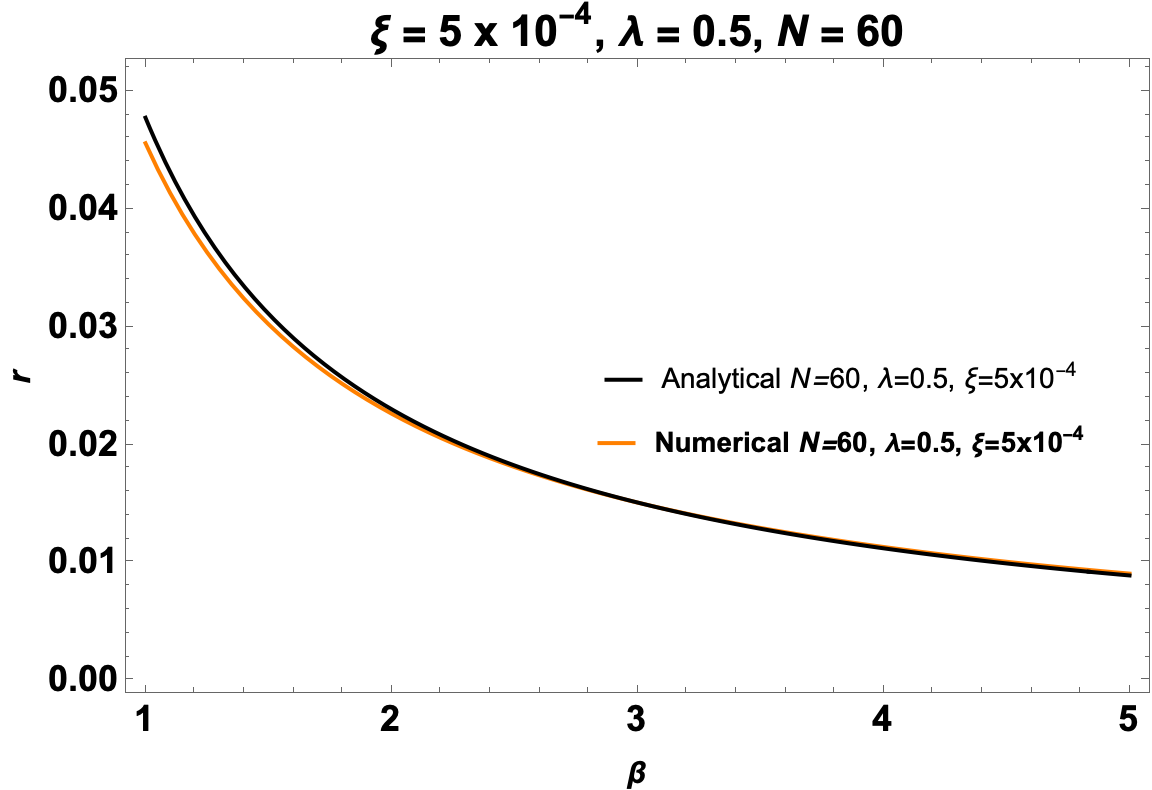}
\includegraphics[width=8 cm]{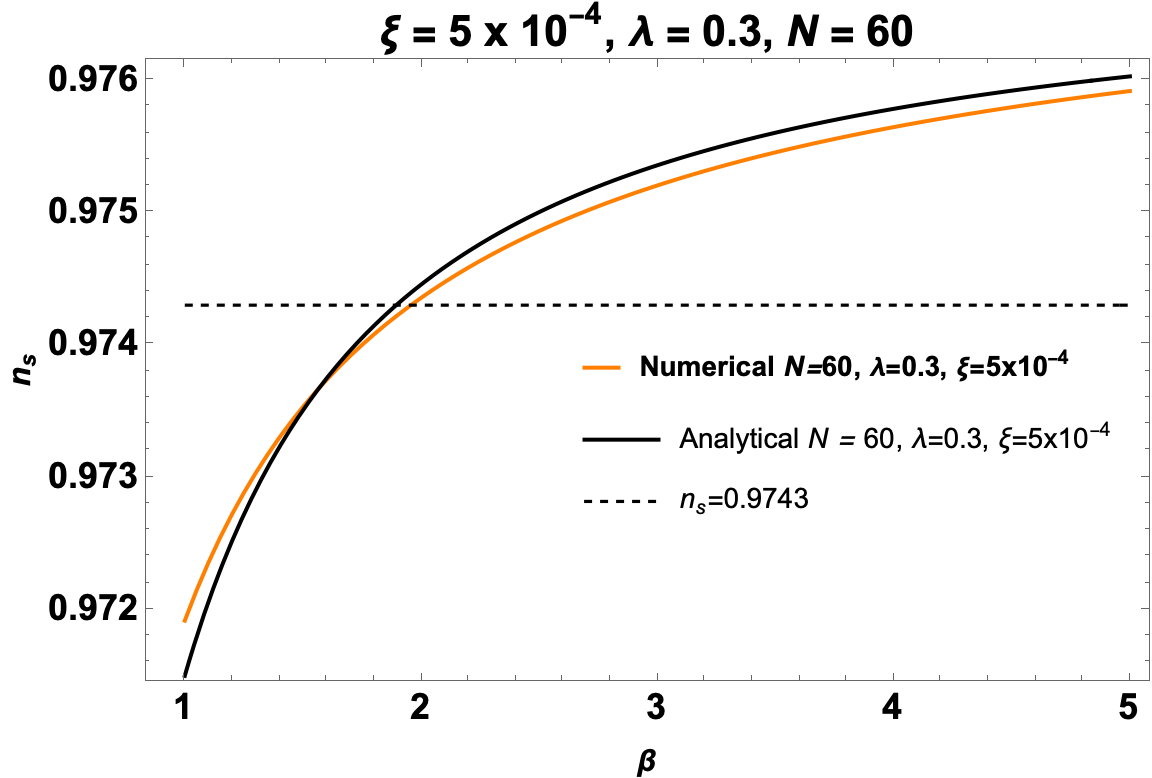}
\includegraphics[width=8 cm]{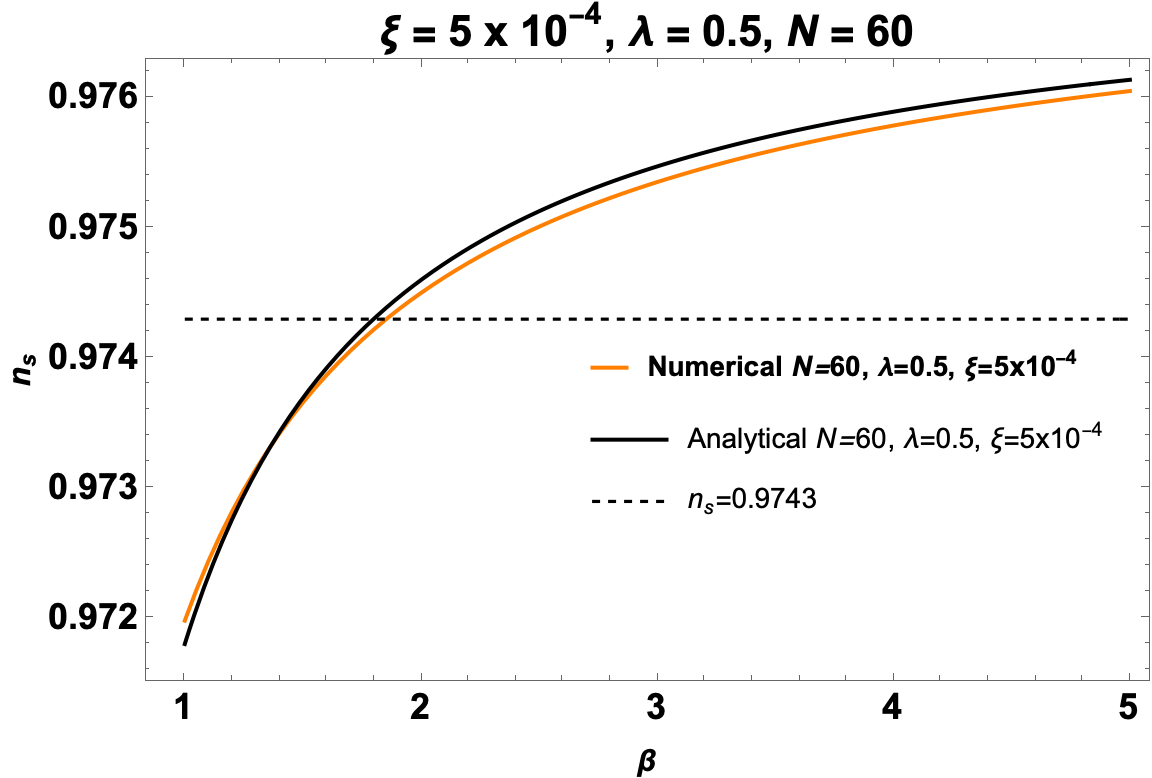}
\caption{We present a comparison between the analytical predictions obtained from the perturbative expansion in the non-minimal coupling parameter $\xi$ and the exact numerical results for the tensor-to-scalar ratio $r$ and the scalar spectral index $n_s$, evaluated at $N=60$ for two representative values of the self-coupling parameter, $\lambda = 0.3$ and $\lambda = 0.5$, with $\xi = 5 \times 10^{-4}$.}\label{numer}
\end{figure}
In Fig.(\ref{numer}), we compare the analytical results derived from a perturbative expansion in the non-minimal coupling parameter $\xi$ with the corresponding exact numerical calculations for the tensor-to-scalar ratio $r$ and the scalar spectral index $n_s$, evaluated at $N=60$ for two representative choices of the self-coupling parameter, $\lambda = 0.3$ and $\lambda = 0.5$, with $\xi = 5 \times 10^{-4}$. The upper panels show the evolution of the tensor-to-scalar ratio $r$ as a function of the parameter $\beta$. In both cases, $r$ decreases monotonically with increasing $\beta$, indicating a progressive suppression of tensor modes in the large-$\beta$ regime. The analytical curves are agreement with the numerical results over the entire range $1 \leq \beta \leq 5$, demonstrating that the perturbative expansion in $\xi$, truncated at leading order, provides a reliable approximation for $r$ when $\xi$ is sufficiently small. 

The lower panels display the corresponding behavior of the scalar spectral index \( n_s \).
As \( \beta \) increases, \( n_s \) gradually rises and asymptotically approaches values
consistent with the Planck 2018 constraint, \( n_s \simeq 0.9743 \), indicated by the
horizontal dashed line. Small deviations between the analytical and numerical curves
are observed at low values of \( \beta \in [1,5] \), where higher-order corrections in
\( \xi \) become more significant. 

Nevertheless, taking the numerical results as
reference values, the averaged percentage differences for \( n_s \) and \( r \) remain
below \( 1\% \) and \( 10\% \), respectively, for \( \lambda = 0.3 \), while for
\( \lambda = 0.5 \) they remain below \( 1\% \) and \( 5\% \), respectively, over the
considered parameter range (see Tables~\ref{tab:non1} and~\ref{tab:non2}). In summary, Fig.~(\ref{rnsnon}) demonstrates that the analytical expressions derived from the perturbative expansion in $\xi$ is consistent with the numerical analysis for the considered parameter space. This consistency validates the use of the analytical framework for investigating the inflationary predictions of the non-minimal model and confirms that leading-order contributions in $\xi$ are sufficient to capture the essential features of both $r$ and $n_s$ in observationally relevant regimes.

\section{Confrontation with ACT Constraints}\label{sec4}
Now we confront the predictions of the tensor-to-scalar ratio $r$ and the scalar spectral index $n_s$ with the ACT data \cite{ACT:2025fju,ACT:2025tim}, the Planck data \cite{Planck:2018jri} and the updated Planck constraints on the tensor-to-scalar ratio \cite{BICEP:2021xfz}. We start with the minimal case, and then the non-minimally-coupled model.
\subsection{Minimally coupled case}
Let us first consider the minimally coupled $\beta$-exponential inflationary scenario and confront its predictions with the latest observational constraints. Using Eq.~(\ref{ns1}), the scalar spectral index $n_s$ depends only on the e-folding number $N$ and the deformation parameter $\beta$, while the tensor-to-scalar ratio $r$ is governed by Eq.~(\ref{17}). These compact relations enable a direct comparison with the most recent ACT and DESI data without relying on additional potential parameters.

Applying the ACT + DESI (P-ACT-LB) limits on the scalar tilt, $n_s = 0.9743 \pm 0.0034$ (68\% C.L.)~\cite{ACT:2025fju,ACT:2025tim}, we find the allowed interval $0.518 < \beta < 3.696$ for $N = 60$ at the $2\sigma$ level. This range indicates that moderately positive values of $\beta$ lead to a scalar spectrum slightly bluer than the universal attractor prediction $n_s = 1 - 2/N \simeq 0.9667$. The $\beta$-exponential potential therefore provides a natural mechanism to lift the spectral index toward the ACT-favoured region while preserving successful slow-roll evolution.

The overall behaviour can be visualized in Fig.~(\ref{rns}), where the theoretical trajectories in the $(r,n_s)$ plane are compared with the Planck 2018 (red) and P-ACT (green) confidence regions. For both $N = 50$ and $N = 60$, the $\beta$-exponential predictions fall within or very near the $2\sigma$ region of the ACT + DESI contour, clearly outperforming the universal attractor relation. Hence, even in the absence of a non-minimal coupling, the $\beta$-exponential inflationary potential provides a simple and phenomenologically viable framework consistent with the latest cosmological observations. Nevertheless, as we show in the next subsection, incorporating a small positive coupling between the inflaton and gravity further improves agreement with the ACT + DESI data by reducing $r$ while keeping $n_s$ nearly unchanged.

\subsection{Non-minimally coupled case}
In the framework of the non-minimally coupled $\beta$-exponential model, we examined whether the introduction of a coupling between the inflaton field and gravity can improve the model’s consistency with present-day cosmological observations. In particular, we investigated how the coupling parameter $\xi$ modifies the inflationary dynamics and affects the inflationary observables, namely the scalar spectral index $n_s$ and the tensor-to-scalar ratio $r$. The predictions of the model were then confronted with the most recent results from the Atacama Cosmology Telescope (ACT) in combination with DESI and Planck data.

\begin{figure}
\includegraphics[width=8 cm]{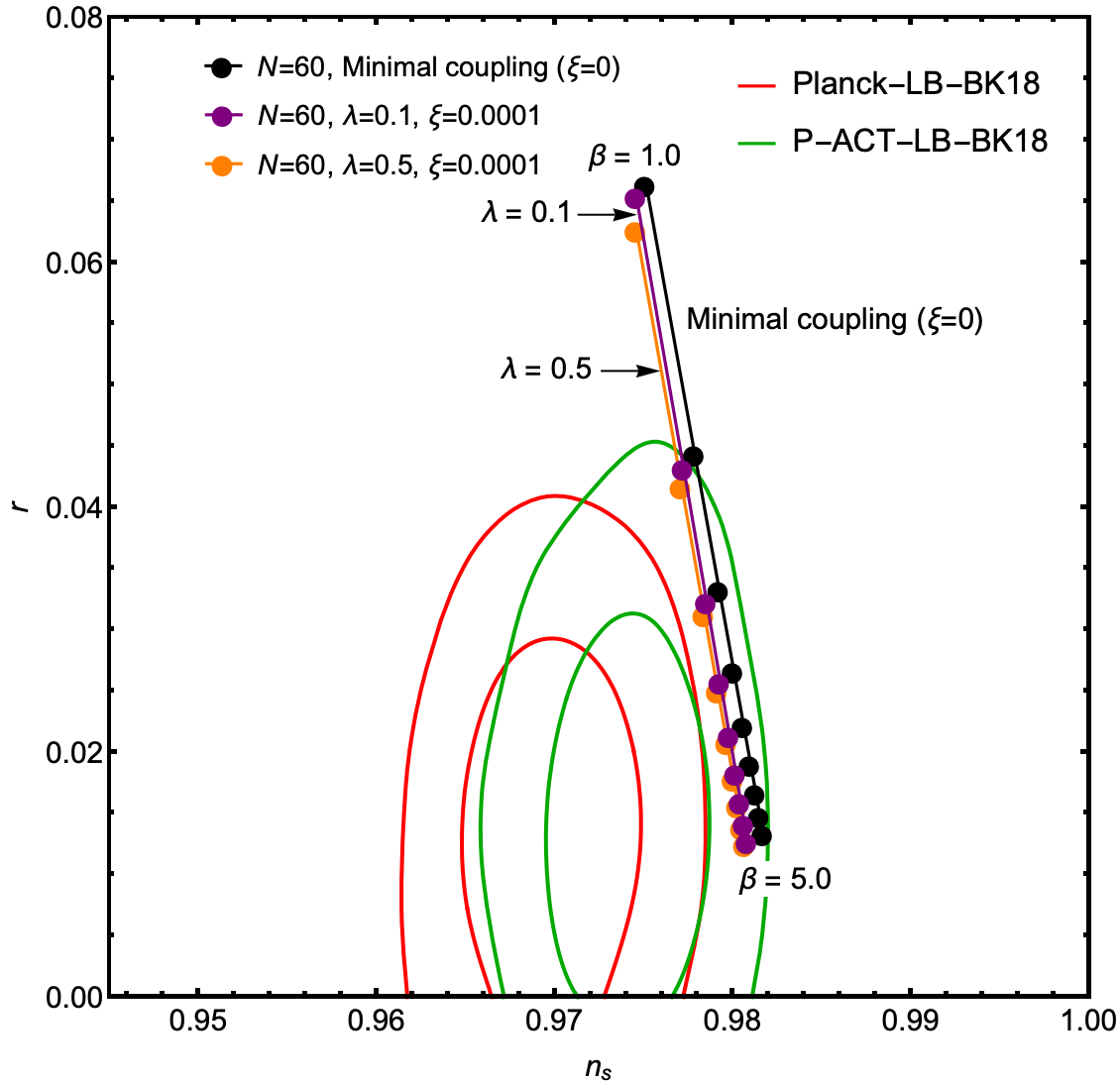}
\includegraphics[width=8 cm]{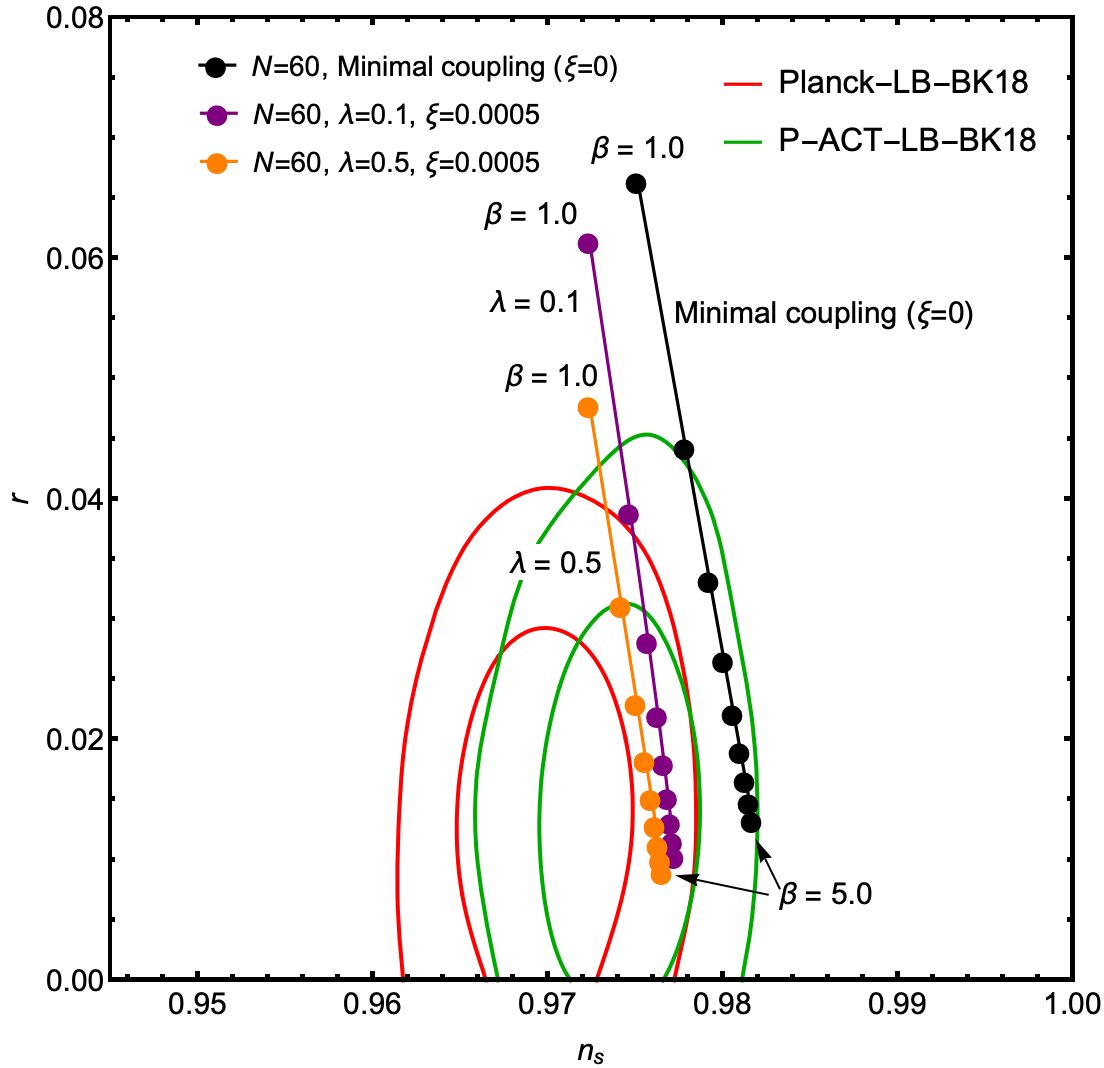}
\caption{Constraints on the scalar and tensor primordial power spectra, shown in the $r-n_{s}$
parameter space predicted by the nonminimally-coupled model (color) compared with a minimally-coupled one (black). The bounds on $r$ are primarily determined by the BK18 observations, whereas the limits on $n_s$ are set by Planck (red) and P-ACT (green) data. We fix $N=60$, $\lambda=0.1$ (purple), $\lambda=0.5$ (orange) and vary a parameter $\beta$ from $1$ to $5$.}\label{rnsnon}
\end{figure}

Figure~(\ref{rnsnon}) illustrates the predictions of the nonminimally coupled inflationary model in the
\( r\text{--}n_s \) plane, compared with the minimally coupled scenario and current observational constraints. The colored trajectories correspond to the nonminimal coupling
cases, while the black curves represent the minimally coupled limit (\( \xi = 0 \)). The observational bounds are shown by the Planck--LB--BK18 (red) and
P-ACT--LB--BK18 (green) confidence contours. For a fixed number of e-folds \( N = 60 \), the model predictions are displayed for two
representative values of the coupling parameter, \( \lambda = 0.1 \) (purple) and
\( \lambda = 0.5 \) (orange), while the parameter \( \beta \) is varied from 1 to 5.
As \( \beta \) increases, the predicted points move toward lower values of the
tensor-to-scalar ratio \( r \) and slightly higher values of the scalar spectral index
\( n_s \). This behavior reflects the progressive flattening of the effective inflationary
potential, which suppresses tensor modes while mildly enhancing scalar perturbations.

The minimally coupled case (black curves) generally predicts slightly larger values of
\( r \) compared to the nonminimally coupled scenarios. Introducing a small nonminimal
coupling shifts the trajectories downward in \( r \), improving compatibility with the
BK18 bounds, while keeping \( n_s \) within the region favored by Planck and ACT
observations. This demonstrates that even a weak nonminimal coupling can play an important
role in reconciling theoretical predictions with observational data.

\begin{figure}
\includegraphics[width=10 cm]{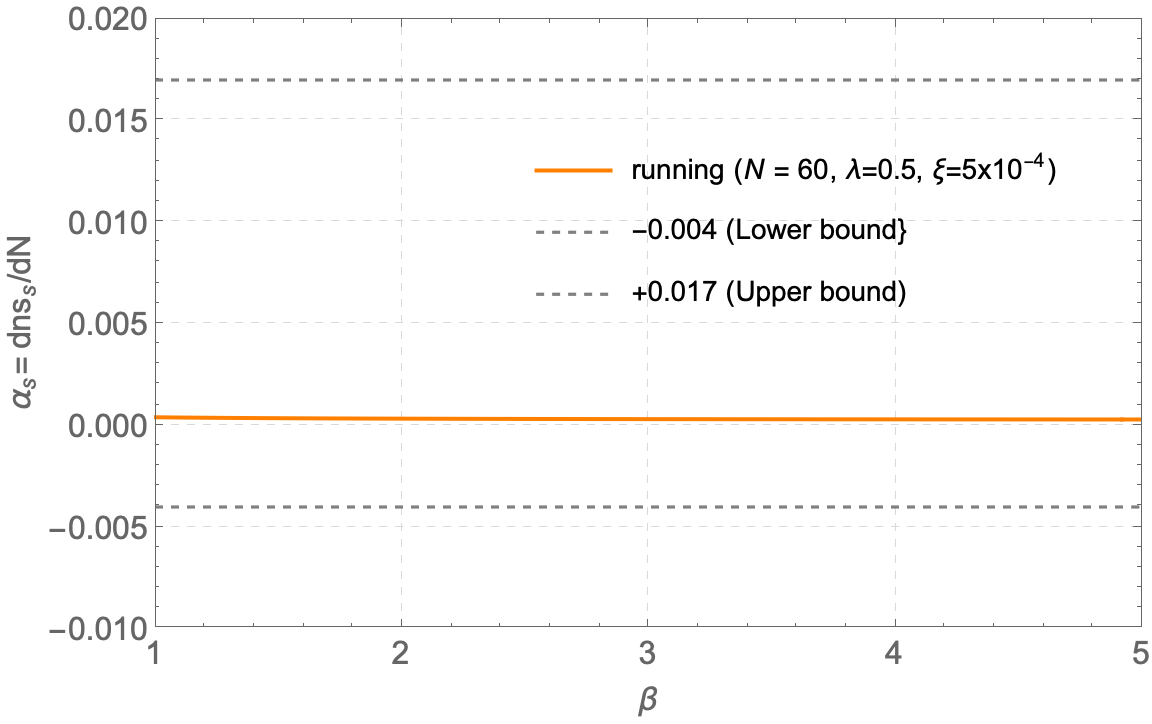}
\caption{The running of the spectral index predicted by our model (solid line) is shown alongside the ACT 95\% confidence limits (dashed lines) over the interval $\beta\in[1,\,5]$ and using $N=60,\,\lambda=0.5,\,\xi=0.0005$.}\label{runss}
\end{figure}

Moreover, for both values of \( \lambda \), the model predictions remain well inside the
combined Planck and ACT confidence regions over the full range
\( \beta \in [1,5] \). In particular, larger values of \( \beta \) lead to predictions
clustered around \( n_s \simeq 0.974-0.976 \) with \( r \lesssim 0.03 \), which lie
comfortably within current observational limits. The relatively low $r$ predicted in this regime implies an inflationary energy scale $V^{1/4} \sim {\cal O}( 10^{-3})M_p$, corresponding to a Hubble rate of order $H \sim 10^{13}\,\mathrm{GeV}$. These values are comparable to those found in Starobinsky-type~\cite{Starobinsky:1980te,Kaiser:1994vs,Bezrukov:2007ep,Bezrukov:2008ej} and $\alpha$-attractor models~\cite{Kallosh:2013tua,Kallosh:2013hoa,Kallosh:2013maa,Kallosh:2013yoa,Kallosh:2014rga,Kallosh:2015lwa,Roest:2015qya,Linde:2016uec,Terada:2016nqg,Ueno:2016dim,Odintsov:2016vzz,Akrami:2017cir,Dimopoulos:2017zvq}, yet the $\beta$-exponential form introduces an additional degree of freedom that shifts the spectral tilt upward without invoking non-canonical kinetic terms or higher-curvature corrections.
There is another relevant observable, the running of the spectral index, which is defined as $\alpha_s = \dfrac{d n_s}{d \ln k}$ with $k$ being the comoving wavenumber of a primordial mode. We can also confront the running of the spectral index for the current model with the ACT data. We can express $\alpha_s$ in the following way
\begin{eqnarray}
\alpha_{s}=\frac{d n_s}{d \ln k}=\frac{d n_s}{d N}\frac{d N}{d \ln k}\,.\label{als}
\end{eqnarray}
However, taking the relation $\dfrac{d N}{d \ln k}=\dfrac{1}{1-\varepsilon}$, the final expression for the running of the spectral index reads
\begin{eqnarray}
\alpha_{s}=\frac{d n_s}{d N}\frac{1}{1-\varepsilon}\,.\label{als1}
\end{eqnarray}
For the present model, we find, up to first-order in $\xi$:
\begin{eqnarray}
\alpha_{s}&=&\frac{d n_s}{d N}\frac{1}{1-\varepsilon}\nonumber\\&\simeq& \frac{8 \beta  (2 \beta +1)}{(4 \beta  N+1)^2}\nonumber\\&&+\frac{4}{3 \beta ^2 \lambda ^3 (4 \beta N+1)^{7/2}}\Bigg(-128 \sqrt{2} \beta ^5 \lambda ^2 N^2+16 \sqrt{2} \beta ^4 \lambda ^2 N (9 N-4)\nonumber\\&&+8 \beta ^3 \lambda ^2 \left(9 \lambda  \sqrt{4 \beta N+1}+\sqrt{2} ((9-2 N) N-1)\right)+4 \sqrt{\beta ^2 \lambda ^4} \sqrt{8 \beta  N+2}\nonumber\\&&+\beta ^2 \left(-64 \sqrt{\beta ^2 \lambda ^4} \sqrt{8 \beta  N+2}+30 \lambda ^3 \sqrt{4 \beta  N+1}+\sqrt{2} \lambda ^2 (9-8 N)\right)\nonumber\\&&-\beta  \left(\sqrt{2} \lambda ^2+24 \sqrt{\beta ^2 \lambda ^4} \sqrt{8 \beta N+2}+3 \lambda ^3 \sqrt{4 \beta N+1}\right)\Bigg)\xi\,.\label{als11}
\end{eqnarray}
The running of the spectral index predicted by our model can now be compared with the ACT observational bounds. As shown in Fig.~(\ref{runss}), the model predictions lie well within the ACT 95\% confidence limits for the running of the spectral index. Our results alongside the ACT 95\% confidence limits place the running in the interval $\alpha_{s}\in[0.0003104,\,0.0005506]$. Given the relevant values of $\beta$ as $1/2 \leq \beta \leq 4$, the $\alpha_{s}$ satisfies the constraint.

In single-field inflationary scenarios, the number of e-folds between horizon exit of the pivot scale and the end of inflation is determined by the post-inflationary reheating dynamics, including the effective equation-of-state parameter during reheating and the reheating temperature. Following standard analyses (see, e.g., Refs.~\cite{Liddle:2003as,Martin:2010kz,Martin:2013tda,Dai:2014jja,Cook:2015vqa}), the number of e-folds can be schematically determined. Within this framework, values in the range $N \simeq 50-60$ naturally arise for physically well-motivated reheating scenarios. In particular, $N \sim 50$ typically corresponds to relatively inefficient or prolonged reheating with an equation of state close to matter domination ($w_{\rm reh} \simeq 0$) and/or a lower reheating temperature, whereas $N \sim 60$ is associated with rapid or nearly instantaneous reheating, or with a stiffer reheating phase ($w_{\rm reh} \gtrsim 1/3$).

In the present work, the reheating phase is not modeled explicitly, and a detailed reheating analysis is therefore beyond the scope of the manuscript. Nevertheless, the choices $N \sim 50$ and $N \sim 60$ are not arbitrary; they are adopted as representative benchmark values spanning the range of e-folds expected from plausible reheating histories. This allows us to assess the robustness of the predicted spectral tilt and tensor-to-scalar ratio against uncertainties associated with the reheating epoch. We have clarified this point in the revised manuscript and indicate that a dedicated study of reheating constraints will be addressed in future work.

\section{Conclusions}\label{Con}
In this work, we have explored the phenomenology of the $\beta$-exponential inflationary model under both minimal and non-minimal couplings between the inflaton field and gravity, motivated by the recent ACT and DESI observations that suggest a slightly higher scalar spectral index $n_s$ than the canonical Planck 2018 value. The $\beta$-exponential potential extends the standard power-law form by introducing the parameter $\beta$, which naturally interpolates between exponential and power-law inflationary behaviors. This generalization provides an additional degree of freedom to accommodate new cosmological data, particularly those from ACT that mildly challenge the predictions of the conventional attractor models.

Our slow-roll analysis demonstrates that the model consistently reproduces viable inflationary observables within the range permitted by current CMB constraints. For the minimally coupled case, the analytical relations between $n_s$, $r$, and the e-fold number $N$ reveal that the predictions remain well within the $2\sigma$ range of Planck and ACT data. Specifically, for $N = 60$ and $\beta = 3.5$, we find $r \approx 0.019$ and $n_s \approx 0.9810$, values that are consistent with the ACT+DESI (P-ACT-LB) upper limits and comfortably below the current bound $r < 0.036$.

The non-minimally coupled extension introduces richer inflationary dynamics and additional parametric control through the coupling constant $\xi$. By employing a perturbative expansion in the small-$\xi$ regime, we derive analytical expressions for the scalar spectral index $n_s$ and the tensor-to-scalar ratio $r$ that smoothly recover the minimally coupled results in the limit $\xi \to 0$. These analytical predictions are further validated by a full numerical analysis of the inflationary dynamics. Our results show that even a very small positive non-minimal coupling ($\xi \sim 5\times 10^{-4}$) is sufficient to improve agreement with the ACT+DESI constraints by significantly suppressing the tensor amplitude while leaving the scalar tilt nearly unchanged. For $N=60$ and representative values $\lambda \sim 0.3$--$0.5$ and $\beta \sim \mathcal{O}(1$--$5)$, the model predicts $n_s \simeq 0.974$--$0.976$ and $r \lesssim 0.03$, consistent with ACT, DESI, and BICEP/Keck bounds. This demonstrates that a mild coupling between the inflaton and gravity can naturally reconcile the $\beta$-exponential model with the ACT-inferred enhancement of the scalar spectral index while maintaining a low tensor-to-scalar ratio. 

Moreover, the relatively low $r$ predicted in this regime implies an inflationary energy scale $V^{1/4} \sim {\cal O}(10^{-3})M_p$, corresponding to a Hubble rate of order $H \sim 10^{13}\,\mathrm{GeV}$. These values are comparable to those found in Starobinsky-type and $\alpha$-attractor models. The predictions in the $r$-$n_s$ plane, illustrated in Figs.~(\ref{rns}) and~(\ref{rnsnon}), further demonstrate that both the minimal and non-minimal $\beta$-exponential models remain viable and consistent with the most stringent CMB observations available to date. From a broader perspective, the $\beta$-exponential inflation scenario provides an appealing bridge between exponential-type inflation and more complex, attractor-like frameworks~\cite{Kallosh:2013tua,Kallosh:2013hoa,Kallosh:2013maa,Kallosh:2013yoa,Kallosh:2014rga,Kallosh:2015lwa,Roest:2015qya,Linde:2016uec,Terada:2016nqg,Ueno:2016dim,Odintsov:2016vzz,Akrami:2017cir,Dimopoulos:2017zvq}, offering a smooth interpolation that may reflect a deeper theoretical structure-potentially arising from braneworld cosmology, see e.g., Ref.~\cite{Santos:2017alg}, or higher-dimensional effective field theories. The addition of the non-minimal coupling naturally connects this model to the class of Higgs and Starobinsky-like inflationary frameworks, which remain among the most observationally favored.

Future research directions include extending this framework to account for radiative and thermal corrections, studying the reheating phase and its impact on the inflationary observables, and exploring potential swampland consistency conditions in this class of models. The upcoming precise cosmic microwave background (CMB) experiments, for instance, CORE \cite{COrE:2011bfs}, AliCPT \cite{Li:2017drr}, LiteBIRD \cite{Matsumura:2013aja}, and CMB-S4 \cite{Abazajian:2019eic} will provide a decisive test for such low-$r$ inflationary scenarios. These next-generation observations will further constrain the allowed parameter space of the $\beta$-exponential potential, offering deeper insights into the fundamental dynamics of the early Universe.

\section*{Acknowledgement}\label{ac}
This work has received funding support from
the NSRF via the Program Management Unit for Human Resources \& Institutional Development, Research and Innovation [grant number B39G680009]


\end{document}